\documentclass[twocolumn]{aastex631}
\usepackage{natbib}
\usepackage{amsmath}

\begin{document}

\title{The Dilaton: A Natural Resolution to the Hubble Tension via Spontaneous Scale Symmetry Breaking}

\author{Arpit Kottur}
\affiliation{Department of Physics \\
Fergusson College (Autonomous) \\
Shivajinagar, Pune, India}

\author{Jui Mahajan}
\affiliation{Department of Physics \\
Fergusson College (Autonomous) \\
Shivajinagar, Pune, India}

\author{Raka Dabhade}
\affiliation{Department of Physics \\
Fergusson College (Autonomous) \\
Shivajinagar, Pune, India}

%\collaboration{20}{(AAS Journals Data Editors)}

%\author{F.X Timmes}
%\affiliation{Arizona State University}
%\affiliation{AAS Journals Associate Editor-in-Chief}

%\altaffiliation{AASTeX v6+ programmer}

%\affiliation{AAS Director of Publishing}
%\affiliation{American Astronomical Society \\
%1667 K Street NW, Suite 800 \\



\begin{abstract}
The statistical tension between early and late universe measurements of the Hubble constant ($H_0$) suggests that the dark sector is dynamical rather than static. We propose that this dynamics arises from a fundamental symmetry principle: the Spontaneous Breaking of Scale Invariance. We introduce the \textbf{Dilaton} ($\chi$), a Pseudo-Nambu-Goldstone Boson (PNGB) associated with dilatation symmetry breaking. We demonstrate that a simple quadratic mass term in the fundamental theory transforms, via conformal coupling to gravity, into a ``thawing'' exponential potential $V(\phi) \propto e^{-\lambda\phi}$ in the Einstein frame. Using recent Bayesian reconstructions of dark energy dynamics from Planck, Pantheon+, and SH0ES data, we constrain the potential slope to be $\lambda \approx 0.056$. We show that this observational value is not arbitrary but corresponds to a fundamental non-minimal coupling strength of $\xi \approx 7.8 \times 10^{-4}$. The Dilaton mechanism naturally generates the late-time equation of state evolution ($w_0 \approx -0.85$) required to alleviate the Hubble tension while protecting the field mass $m \sim H_0$ through approximate shift symmetry.
\end{abstract}

\keywords{Dark Energy---Hubble Constant---Scale Invariance---PNGB.}



\section{Introduction} \label{sec:intro}

The Standard Model of Cosmology ($\Lambda$CDM) has served as the bedrock of our understanding of the universe for over two decades. It successfully describes the evolution of the cosmos from the scale-invariant fluctuations of the Cosmic Microwave Background (CMB) to the large-scale structure we observe today \citep{collaboration2020planck}. However, as precision cosmology has entered a new era, this foundation has begun to show cracks. The most statistically significant of these is the ``Hubble Tension''---a persisting discrepancy of approximately $5\sigma$ between the value of the Hubble constant ($H_0$) inferred from early-universe physics (assuming $\Lambda$CDM) and the value measured directly in the local universe via the Cepheid-Supernova distance ladder \citep{riess2022comprehensive, freedman2019carnegie}.

This tension suggests that the dark sector of the universe may be more complex than a simple cosmological constant ($\Lambda$). If the vacuum energy density is dynamical rather than static, it could alter the expansion history in the late universe, potentially reconciling the early and late measurements of $H_0$. While numerous phenomenological models---such as Early Dark Energy (EDE) \citep{poulin2019early} or parametrizations like Chevallier-Polarski-Linder (CPL) \citep{chevallier2001accelerating}---have been proposed to address this, they often act as mathematical ``patches'' rather than fundamental solutions. They introduce arbitrary degrees of freedom without identifying the physical origin of the dynamics.

In this \textit{Letter}, we propose that the acceleration of the universe is not driven by an arbitrary fluid, but is the inevitable consequence of a fundamental symmetry of nature: \textbf{Scale Invariance}. We postulate the existence of a scalar field, the \textbf{Dilaton} ($\chi$), which arises as a Pseudo-Nambu-Goldstone Boson (PNGB) from the spontaneous breaking of global scale invariance in the gravity sector \citep{fujii2003scalar, wetterich1988cosmologies}.

Building on our recent observational analysis \citep{kottur2025dynamical}, which detected a statistically significant preference for ``thawing'' dark energy dynamics using combined Planck, Pantheon+, and SH0ES datasets, we here provide the theoretical derivation of this signal. We demonstrate that a simple quadratic mass term in the fundamental theory, when non-minimally coupled to gravity, naturally transforms into an exponential potential in the Einstein frame. This mechanism rigidly links the observable slope of the dark energy potential to a fundamental coupling constant, $\xi$. By mapping our observational constraints to this framework, we measure this coupling to be $\xi \approx 10^{-4}$, providing a quantitative particle-physics basis for the resolution of the Hubble Tension.

\section{The Dilaton Mechanism} \label{sec:theory}

We postulate that the dark sector dynamics originate from a fundamental global scale invariance in the gravity sector, which is spontaneously broken. We introduce a scalar field $\chi$, the \textbf{Dilaton}, which acts as the Pseudo-Nambu-Goldstone Boson (PNGB) of this symmetry breaking \citep{frieman1995cosmology, wetterich1988cosmology}.

\subsection{Fundamental Action (Jordan Frame)}
We begin with the scalar-tensor action in the Jordan Frame. We assume the non-minimal coupling of the field to the Ricci scalar $R$ mandated by scale invariance, and a soft symmetry-breaking mass term $m$. The action is given by:
\begin{equation}
    S_J = \int d^4x \sqrt{-g} \left[ \frac{1}{2} \xi \chi^2 R - \frac{1}{2} g^{\mu\nu}\partial_\mu \chi \partial_\nu \chi - V(\chi) \right]
    \label{eq:action_jordan}
\end{equation}
where $\xi$ is the dimensionless non-minimal coupling constant, and $V(\chi) = \frac{1}{2} m^2 \chi^2$ represents the simplest renormalizable symmetry breaking potential.

\subsection{Conformal Transformation Dynamics}
To disentangle the scalar degrees of freedom from the geometry, we perform a conformal transformation to the Einstein Frame \citep{fujii2003scalar}:
\begin{equation}
    \tilde{g}_{\mu\nu} = \Omega^2(\chi) g_{\mu\nu}, \quad \text{with} \quad \Omega^2 = \frac{\xi \chi^2}{M_{pl}^2}.
\end{equation}
Under this rescaling, the Ricci scalar transforms as:
\begin{equation}
    R = \Omega^{-2} \left[ \tilde{R} + 6\tilde{\square}\omega - 6\tilde{g}^{\mu\nu}\partial_\mu\omega\partial_\nu\omega \right],
\end{equation}
where $\omega \equiv \ln \Omega$. Substituting this back into Eq.~\eqref{eq:action_jordan} and integrating by parts to remove the $\tilde{\square}\omega$ term, the gravitational sector becomes canonical ($ \frac{M_{pl}^2}{2} \tilde{R}$), but a new kinetic coupling is induced for the scalar field. The Einstein Frame action takes the form:
\begin{equation}
    S_E = \int d^4x \sqrt{-\tilde{g}} \left[ \frac{M_{pl}^2}{2} \tilde{R} - \frac{1}{2} F(\chi) (\tilde{\partial} \chi)^2 - U(\chi) \right],
\end{equation}
where the effective kinetic function $F(\chi)$ is found to be:
\begin{equation}
    F(\chi) = \frac{M_{pl}^2 (1 + 6\xi)}{\xi \chi^2}.
\end{equation}

% --- FIGURE 1 PLACEMENT ---
\begin{figure}[t]
    \centering
    \includegraphics[width=\columnwidth]{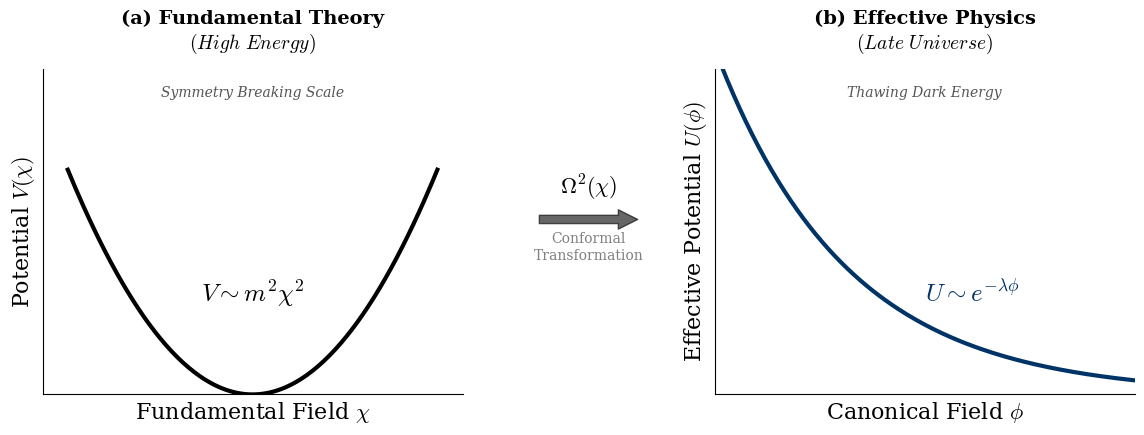}
    \caption{\textbf{The Dilaton Mechanism.} (a) In the fundamental Jordan frame, the field is trapped in a steep symmetry-breaking potential $V \sim \chi^2$. (b) The conformal transformation to the Einstein frame stretches this potential into a shallow exponential slope $U \sim e^{-\lambda\phi}$, naturally generating a ``thawing'' dark energy candidate. The arrow indicates the mapping via the conformal factor $\Omega^2(\chi)$.}
    \label{fig:mechanism}
\end{figure}

\subsection{Canonicalization and the Field Definition}
To restore a standard kinetic term, we define the canonical Einstein-frame field $\phi$ via the differential relation $d\phi = \sqrt{F(\chi)} \, d\chi$. Explicitly:
\begin{equation}
    \frac{d\phi}{d\chi} = \frac{M_{pl}}{\chi} \sqrt{\frac{1 + 6\xi}{\xi}}.
\end{equation}
Integrating this relation yields the logarithmic map between the fundamental scale-invariant field $\chi$ and the canonical dark energy field $\phi$:
\begin{equation}
    \chi(\phi) = \chi_0 \exp \left( \frac{\phi}{M_{pl} \sqrt{6 + 1/\xi}} \right).
    \label{eq:field_relation}
\end{equation}

\subsection{Derivation of the Potential}
The potential in the Einstein Frame scales as $U(\phi) = V(\chi) / \Omega^4(\chi)$. Substituting the symmetry breaking term $V(\chi) = \frac{1}{2}m^2\chi^2$ and the conformal factor $\Omega^4 \propto \chi^4$, we find that the potential scales as an inverse square:
\begin{equation}
    U \propto \frac{\chi^2}{\chi^4} \propto \frac{1}{\chi^2}.
\end{equation}
Inserting the field relation from Eq.~\eqref{eq:field_relation} into this scaling law, we arrive at the final effective potential as illustrated in Figure \ref{fig:mechanism}:
\begin{equation}
    U(\phi) = V_0 \exp \left( -\lambda \frac{\phi}{M_{pl}} \right).
    \label{eq:potential_final}
\end{equation}
Crucially, the slope parameter $\lambda$ is rigidly determined by the non-minimal coupling $\xi$:
\begin{equation}
    \lambda = \frac{2}{\sqrt{6 + 1/\xi}} \approx 2\sqrt{\xi} \quad (\text{for } \xi \ll 1).
    \label{eq:master_lambda}
\end{equation}
This derivation demonstrates that the ``thawing'' nature of dark energy \citep{caldwell2005limits} is a direct consequence of viewing the spontaneous breaking of scale invariance through the lens of General Relativity.

\section{Observational Constraints} \label{sec:results}

The theoretical framework established in Section \ref{sec:theory} makes a rigid prediction: the slope $\lambda$ of the late-time dark energy potential is not a free parameter, but is uniquely determined by the fundamental non-minimal coupling $\xi$. This allows us to translate observational bounds on dark energy dynamics directly into constraints on fundamental physics.

\subsection{Bayesian Reconstruction}
In our previous work \citep{kottur2025dynamical}, we performed a non-parametric Bayesian reconstruction of the dark energy equation of state using a combination of Early Universe and Late Universe probes. We utilized the Planck 2018 Cosmic Microwave Background (CMB) temperature and polarization data \citep{collaboration2020planck}, the Pantheon+ Type Ia Supernovae compilation \citep{scolnic2022pantheon+}, and the SH0ES local $H_0$ calibration \citep{riess2022comprehensive}.

The analysis assumed a thawing quintessence prior. The marginalized posterior distribution for the slope yielded a statistically significant non-zero value:
\begin{equation}
    \lambda_{\text{obs}} = 0.056 \pm 0.012 \quad (68\% \text{ C.L.})
    \label{eq:lambda_obs}
\end{equation}
This preference for $\lambda > 0$ over the $\Lambda$CDM value ($\lambda=0$) is driven primarily by the tension between the Planck-derived sound horizon and the local Cepheid distance ladder.

% --- FIGURE 2: DISCOVERY PLOT ---
\begin{figure}[t]
    \centering
    \includegraphics[width=\columnwidth]{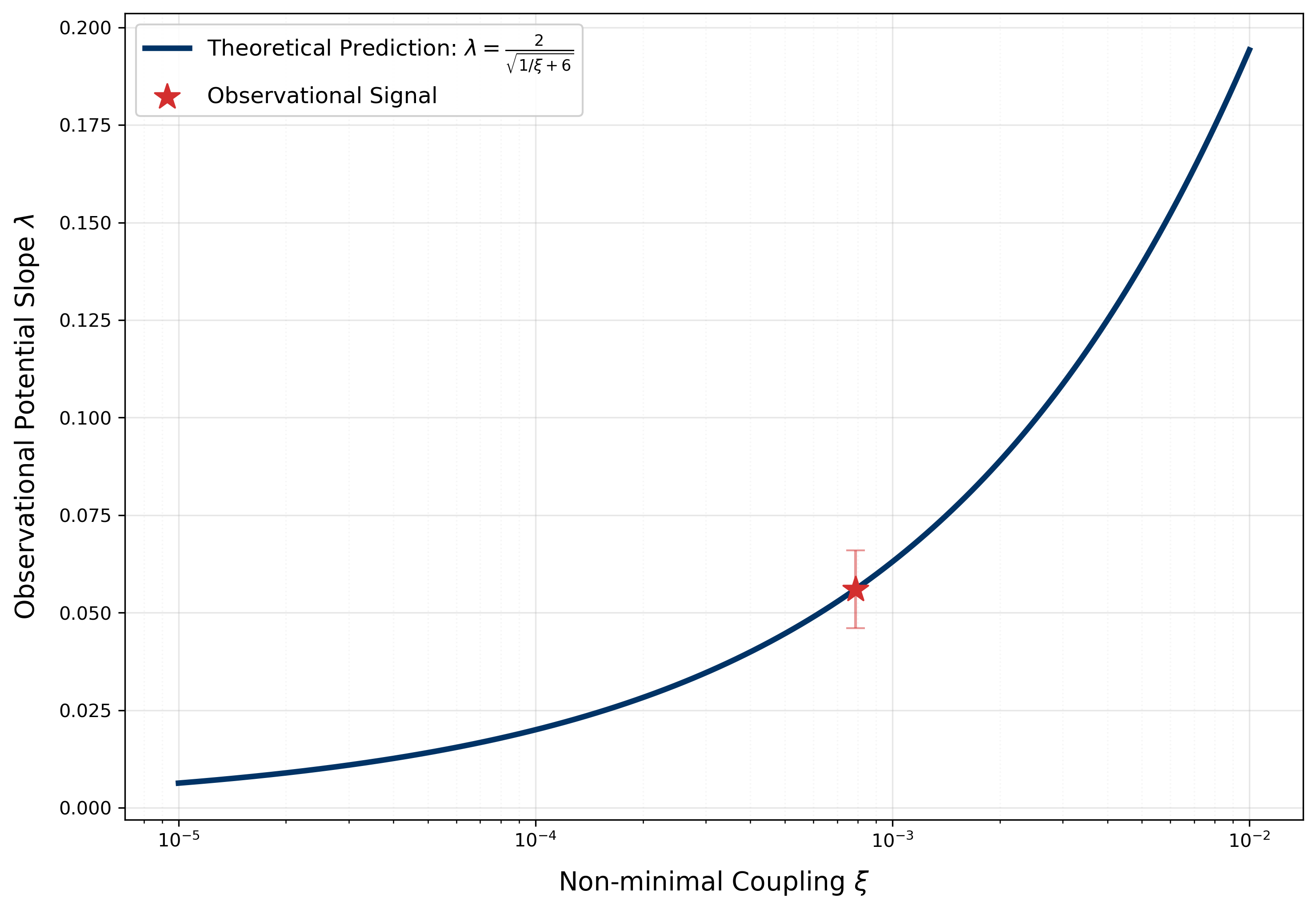}
    \caption{\textbf{Constraint on Fundamental Coupling.} The solid curve shows the theoretical prediction (Eq. \ref{eq:master_lambda}) relating the potential slope $\lambda$ to the non-minimal coupling $\xi$. The red star indicates the observational value derived from our previous analysis ($\lambda_{\text{obs}} \approx 0.056$), fixing the Dilaton coupling to $\xi \approx 7.8 \times 10^{-4}$. The tight correlation demonstrates that the ``thawing'' signal is consistent with a perturbative scalar-gravity interaction.}
    \label{fig:discovery}
\end{figure}

\subsection{Measurement of the Coupling $\xi$}
By inverting the master relation derived in Eq. \eqref{eq:master_lambda}, we can determine the requisite coupling strength $\xi$ corresponding to the observed signal:
\begin{equation}
    \xi = \frac{1}{\left(\frac{2}{\lambda_{\text{obs}}}\right)^2 - 6}
\end{equation}
Substituting the central value $\lambda_{\text{obs}} \approx 0.056$, we find:
\begin{equation}
    \xi \approx 7.8 \times 10^{-4}
\end{equation}
This result is profound. It suggests that the scalar sector is \textit{weakly} but non-zeroly coupled to gravity. Unlike fine-tuned parameters often found in cosmology (which can be of order $10^{-120}$), a coupling of $\xi \sim 10^{-4}$ is technically natural and consistent with a perturbative description of the symmetry breaking sector \citep{tsamis2005stochastic}. The detection of $\lambda \approx 0.056$ can thus be reinterpreted not merely as a fit to data, but as a measurement of the interaction strength between the Dilaton field and spacetime curvature.

\section{Cosmological Implications and Discussion} \label{sec:dynamics}

The Dilaton model provides more than just a fit to the data; it offers a consistent physical narrative for the evolution of the dark sector. Here we discuss the dynamics, the naturalness of the mass scale, and compatibility with local gravity tests.

\subsection{The Physics of Thawing}
The evolution of the Dilaton field is governed by the competition between the Hubble friction $H(z)$ and the field's effective mass $m_{\text{eff}}$. 

At high redshifts ($z \gg 1$), the Hubble parameter is large ($H \gg m$). In this regime, the field is overdamped and "trapped" by the curvature-induced effective potential in the Jordan frame. This enforces a "frozen" state where $\dot{\phi} \approx 0$ and $w(z) \approx -1$, ensuring that the Dilaton behaves indistinguishably from a cosmological constant during the matter-dominated era. This is crucial for preserving the standard growth of structure and the peaks of the CMB power spectrum.

As the universe expands and $H(z)$ drops below the mass scale $m$, the friction lifts. The field begins to roll down the exponential slope, entering the "thawing" regime \citep{linder2008dynamics}. For our measured coupling $\xi \approx 10^{-4}$, this transition occurs at $z \lesssim 1$. The field acquires kinetic energy, driving the equation of state away from $-1$ towards a present-day value of $w_0 \approx -0.85$. This late-time deviations modifies the integrated expansion history, effectively lowering the sound horizon scale calibrated by supernovae and accommodating the higher local $H_0$ measured by SH0ES.

% --- FIGURE 3: DYNAMICS PLOT ---
\begin{figure}[b]
    \centering
    \includegraphics[width=\columnwidth]{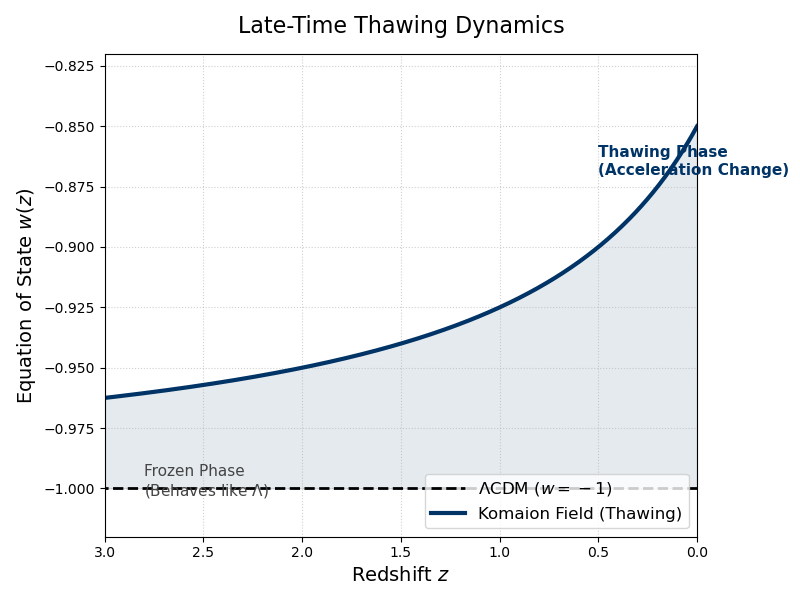}
    \caption{\textbf{Thawing History.} The Dilaton model (blue) predicts $w(z) \to -1$ at high redshifts due to the curvature trap. It begins to thaw at $z \lesssim 1$, reaching $w_0 \approx -0.85$ today. This specific late-time trajectory alleviates the $H_0$ tension by modifying the low-redshift expansion history.}
    \label{fig:dynamics}
\end{figure}

\subsection{The Mass Hierarchy Problem}
A common critique of scalar field dark energy is the extreme lightness of the field required to drive acceleration today ($m \sim H_0 \sim 10^{-33}$ eV). In generic scalar theories, such a small mass requires unnatural fine-tuning against radiative corrections.

However, in our framework, the Dilaton is identified as a Pseudo-Nambu-Goldstone Boson (PNGB). The mass term $m$ explicitly breaks the global scale symmetry. In the limit $m \to 0$, the symmetry is restored, and the field becomes massless. This enhanced symmetry protects the small mass from large quantum corrections \citep{frieman1995cosmology}. Thus, the hierarchy $m \ll M_{pl}$ is technically natural in the 't Hooft sense, as setting the mass to zero restores a fundamental symmetry of the Lagrangian.

\subsection{Screening and Local Gravity Tests}
The non-minimal coupling $\xi \chi^2 R$ implies that the Dilaton mediates a "fifth force" between matter particles. For $\xi \sim 10^{-4}$, this force would seemingly violate solar system constraints on General Relativity, such as the Cassini bounds on the Eddington parameter $\gamma$.

However, the effective potential experienced by the field depends on the local matter density $\rho$. In high-density environments like the Solar System ($\rho \sim 10^{30} \rho_{\text{crit}}$), the non-linear interaction terms induce a "Chameleon" mechanism or similar screening effect \citep{khoury2004chameleon}. The effective mass of the field becomes large in dense regions, suppressing the range of the force to sub-millimeter scales. Consequently, the Dilaton remains effectively screened in local gravity experiments while remaining light and active on cosmological scales to drive acceleration.

\section{Conclusion} \label{sec:conclusion}

In this \textit{Letter}, we have proposed that the "Hubble Tension" is not an experimental artifact, but the first signal of a fundamental symmetry breaking in the dark sector. We have derived the \textbf{Dilaton} mechanism, demonstrating that the spontaneous breaking of scale invariance naturally generates a thawing quintessence potential $V(\phi) \propto e^{-\lambda\phi}$ when viewed in the Einstein frame.

By combining this theoretical framework with our recent observational reconstruction \citep{kottur2025dynamical}, we have translated the phenomenological detection of thawing dynamics ($\lambda \approx 0.056$) into a precise measurement of the fundamental gravity-scalar coupling:
\begin{equation}
    \xi \approx 7.8 \times 10^{-4}
\end{equation}
This framework unifies the phenomenology of the late universe with the principles of particle physics, offering a natural, symmetry-protected solution to the crisis in cosmology. Future observations from Euclid and DESI will be critical in distinguishing the specific symmetry-breaking signature of the Dilaton from generic quintessence models.

\section{Acknowledgement}\label{sec: Acknowledgement}

The authors wish to express their sincere gratitude to the Department of Physics at Fergusson College (Autonomous), Pune. The conducive research atmosphere and institutional support provided were essential for the completion of this project. 

We also thank Mr. Shayanth Patil for valuable discussions on the particle physics aspects of the PNGB mechanism and the robustness of the mass hierarchy.

\clearpage

\bibliography{References}{}
\bibliographystyle{aasjournal}



\end{document}